\documentclass[preprint2]{aastex631}

\usepackage{graphicx}
\usepackage[flushleft]{threeparttable}
\usepackage{amsmath}	
\usepackage{amssymb}	
\usepackage{xcolor}

\newcommand{\s}{\,{\rm s}}      
\newcommand{\ps}{\,{\rm s}^{-1}}

\newcommand{\cm}{\,{\rm cm}}    
\newcommand{\km}{\,{\rm km}}
\newcommand{\kms}{$\km\ps$}

\newcommand{\K}{\,{\rm K}}

\newcommand{\snr}{SN~1006}
\newcommand{\src}{SN~1006~SW}

\newcommand{\RAdot}[4]{{#1}^{{\rm h}}{#2}^{{\rm m}}{#3}\fs{#4}}
\newcommand{\decldot}[4]{{#1}^{\circ}{#2}'{#3}\farcs{#4}}

\newcommand{\decl}[3]{{#1}^{\circ}{#2}'{#3}''}

\begin{document}

\title{X-ray polarization in SN 1006 southwest shows spatial variations and differences with the radio band}

\correspondingauthor{Ping Zhou}
\email{pingzhou@nju.edu.cn}

\author[0000-0002-5683-822X]{Ping Zhou}
\affiliation{School of Astronomy \& Space Science, Nanjing University, 163 Xianlin Avenue, Nanjing 210023, China}
\affiliation{Key Laboratory of Modern Astronomy and Astrophysics, Nanjing University, Ministry of Education, Nanjing 210023, China}

\author[0000-0002-6986-6756]{Patrick Slane}
\affiliation{Center for Astrophysics | Harvard \& Smithsonian, 60 Garden St, Cambridge, MA 02138, USA}

\author[0000-0001-6511-4330]{Dmitry Prokhorov}
\affiliation{Fakult{\"a}t f{\"u}r Physik und Astronomie,  Julius-Maximilians-Universit{\"a}t W{\"u}rzburg, Emil-Fischer-Str. 31, 97074 W{\"u}rzburg, Germany}

\author[0000-0002-4708-4219]{Jacco Vink}
\affiliation{Anton Pannekoek Institute for Astronomy \& GRAPPA, University of Amsterdam, Science Park 904, 1098 XH Amsterdam, The Netherlands}
\author[0000-0003-1074-8605]{Riccardo Ferrazzoli}
\affiliation{INAF Istituto di Astrofisica e Planetologia Spaziali, Via del Fosso del Cavaliere 100, 00133 Roma, Italy}
\author[0000-0001-7363-6489]{William Cotton}
\affiliation{National Radio Astronomy Observatory, 520 Edgemont Road, Charlottesville, VA 22903, USA}
\affiliation{South African Radio Astronomy Observatory, 2 Fir Street, Cape Town, 7925, South Africa}
\author[0000-0002-8848-1392]{Niccolò Bucciantini}
\affiliation{INAF Osservatorio Astrofisico di Arcetri, Largo Enrico Fermi 5, 50125 Firenze, Italy}
\affiliation{Dipartimento di Fisica e Astronomia, Università degli Studi di Firenze, Via Sansone 1, 50019 Sesto Fiorentino (FI), Italy}
\affiliation{Istituto Nazionale di Fisica Nucleare, Sezione di Firenze, Via Sansone 1, 50019 Sesto Fiorentino (FI), Italy}
\author[0000-0001-9108-573X]{Yi-Jung Yang}
\affiliation{Graduate Institute of Astronomy, National Central University, 300 Zhongda Road, Zhongli, Taoyuan 32001, Taiwan}
\affiliation{Laboratory for Space Research, The University of Hong Kong, Cyberport 4, Hong Kong}
\author[0000-0002-8665-0105]{Stefano Silvestri}
\affiliation{Istituto Nazionale di Fisica Nucleare, Sezione di Pisa, Largo B. Pontecorvo 3, 56127 Pisa, Italy}
\author[0000-0002-2954-4461]{Douglas A. Swartz}
\affiliation{Science and Technology Institute, Universities Space Research Association, Huntsville, AL 35805, USA}
\author[0000-0002-3638-0637]{Philip Kaaret}
\affiliation{NASA Marshall Space Flight Center, Huntsville, AL 35812, USA}
\author[0000-0003-4925-8523]{Enrico Costa}
\affiliation{INAF Istituto di Astrofisica e Planetologia Spaziali, Via del Fosso del Cavaliere 100, 00133 Roma, Italy}
\author[0000-0002-5847-2612]{C.-Y. Ng}
\affiliation{Department of Physics, The University of Hong Kong, Pokfulam, Hong Kong}
\author[0000-0002-1875-7701]{Estela Reynoso}
\affiliation{Instituto de Astronom\'{i}a y F\'{i}sica del Espacio (IAFE), Av. Int. Guiraldes 2620, Pabell\'{o}n IAFE, Ciudad Universitaria, 1428 Ciudad Aut\'{o}noma de Buenos Aires, Argentina}
\author[0000-0002-7781-4104]{Paolo Soffitta}
\affiliation{INAF Istituto di Astrofisica e Planetologia Spaziali, Via del Fosso del Cavaliere 100, 00133 Roma, Italy}
\author[0000-0002-9443-6774]{Allyn F. Tennant}
\affiliation{NASA Marshall Space Flight Center, Huntsville, AL 35812, USA}
\author[0009-0002-4427-6976]{Wenlang He}
\affiliation{School of Astronomy \& Space Science, Nanjing University, 163 Xianlin Avenue, Nanjing 210023, China}
\author{David Moffett}
\affiliation{Department of Physics, Furman University, Greenville, SC 29613, USA}

\begin{abstract}

We report the detection of a spatial variation of X-ray polarization in the southwestern shell of \snr\ (\src) using IXPE. 
The shell has an average X-ray polarization degree (PD) of $21.6\%\pm 4.5\%$ and polarization angle (PA) of $-48^\circ \pm 5^\circ$ in the 2--4 keV energy band, similar to those in the northeastern shell. 
The PD varies along \src, with a peak PD$= 40\%\pm 8\%$ in the south and a significantly lower PD $\lesssim 27\%$ (99\% upper limit) in the west where the shell has been proposed to be interacting with an interstellar cloud. The correlation between the PD, which reflects the magnetic orderliness, and the preshock density provides observational evidence that magnetic turbulence and amplification are environment-dependent. 
The high PD detected in the southern region of the shell constrains the magnetic turbulence scale of $\lesssim 0.1$~pc.
Moreover, by comparing the IXPE X-ray and MeerKAT radio polarization measurements for \src, we found that magnetic fields traced by X-ray polarization are nearly radially distributed, whereas those traced by radio polarization tend to follow a direction parallel to the Galactic plane.
This suggests that the X-ray polarization probes freshly amplified magnetic fields from small-scale structures in the immediate postshock region, while the radio traces more extended regions influenced by the pre-existing ambient magnetic fields. 

\end{abstract}

\keywords{Supernova remnants (1667); Polarimetry (1278); Shocks (2086); Cosmic rays (329)}

\section{Introduction} \label{sec:intro}

Supernova remnants (SNRs) influence almost every component of the interstellar medium (ISM), which consists of gas, dust, cosmic rays (CRs), magnetic fields, and radiation \citep{draine11}. While driving shocks into the gas, they accelerate particles to the relativistic regimes (a.k.a.\ CRs) and strongly modify magnetic fields.
Shocks, magnetic fields, and CR acceleration processes are closely coupled --- we cannot fully understand one process without considering how it interacts with the others.
A crucial open question is whether SNRs can accelerate CRs to the PeV energy regime.
For SNRs to behave as PeVatrons, the magnetic field needs to be highly amplified to keep CRs being accelerated near the shock \citep[see a review by][]{blasi13}.

X-ray observations have demonstrated that magnetic fields can be amplified by a factor of 10--100 in young SNRs \citep{vink03a,parizot06,uchiyama07}. However, the mechanism behind magnetic field amplification in SNRs remains poorly understood. Potential mechanisms involve CR-driven instability \citep{bell04} and turbulent dynamo processes \citep[e.g.,][]{inoue13,xu17}. 

Polarization measurements unveil the magnetic field orientation and turbulence level through the electric vector polarization angle (PA, perpendicular to the magnetic field orientation) and polarization degree (PD), respectively, and thus provide insight into this question.
Radio polarimetry of SNRs has been conducted for decades \citep[see references within the review by ][]{dubner15}, whereas X-ray imaging polarization became possible only recently, starting in 2021 with the launch of the Imaging X-ray Polarimetry Explorer \citep[IXPE;][]{weisskopf21}.

IXPE has observed six SNRs to date, providing new knowledge of magnetic fields close to the shock front of young SNRs, some discoveries of which were unexpected \citep[see a recent review by][]{slane24}. Three SNRs observed in the first year of IXPE are Cas~A, Tycho, and SN~1006 (G327.6+14.6, NE shell). They show different PD in the rim from $4.5\%\pm 1.0\%$, $12\%\pm 2\%$ to $22.4\%\pm 3.5\%$ \citep{vink22,ferrazzoli23,zhou2024}.
Although all three of these young SNRs reveal X-ray PA tangential to the shell (corresponding to radially oriented magnetic fields), later IXPE observations of RX~J1713.3$-$3946 \citep{ferrazzoli24} and Vela Jr.\ \citep{prokhorov24} unexpectedly revealed X-ray PAs perpendicular to the shell,
supporting a compression origin of the magnetic fields even for these dynamically young SNRs. Another interesting finding seen so far is that none of these SNRs share similar PD. The reason is unclear but might be related to the environmental density and magnetic field, shock velocity, SNR age, and whether IXPE can resolve the magnetic turbulence scale \citep{slane24}. Other possibilities, such as the ion-neutral damping when CR currents interact with neutrals, might also change the magnetic turbulence level and, thus, the PD \citep{drury96}.

\begin{figure}
\centering
	\includegraphics[width=0.5\textwidth]{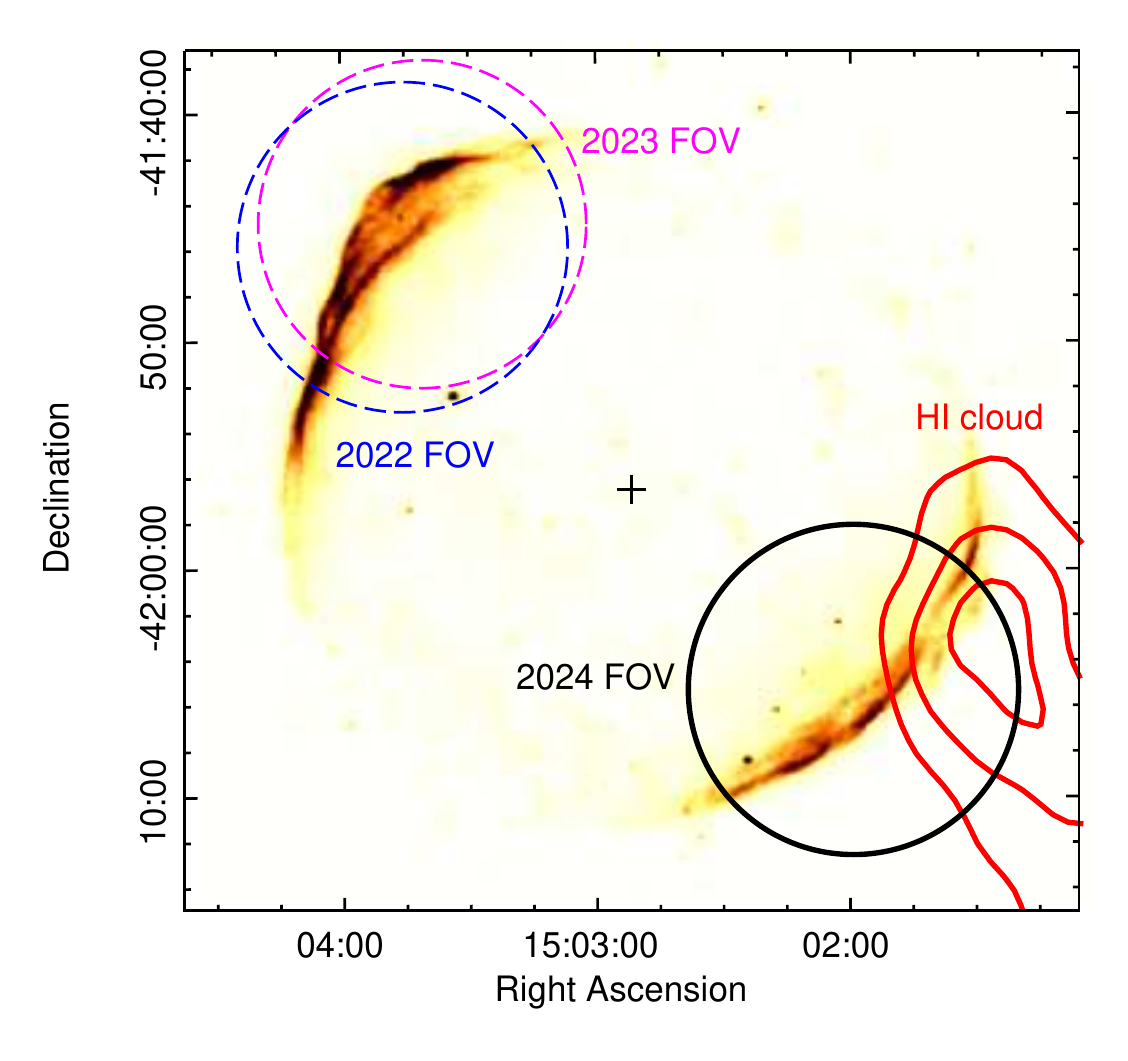}
\caption{Field-of-view (FOV) of the IXPE observations overlaid on the XMM-Newton 2--8~keV image of SN~1006 \citep{li16}. The solid and dashed circles denote the FOVs for the current observation of \src\ and previous observations of SN~1006~NE. The red contours show the HI intensity integrated within the local standard of rest (LSR) velocity interval of 5.7-- 10.7~\kms. The plus sign shows the SNR center \citep[$\RAdot{15}{02}{51}{7}$, $\decl{-41}{56}{33}$, J2000,][]{reynolds86,winkler97}.
}
    \label{fig:fov}
\end{figure}

To examine the influence of ambient density on magnetic turbulence, a viable way is to perform imaging polarimetry of an individual SNR and measure how the polarization properties change as ambient density varies along the rim. 
This motivated us to observe the southwestern shell of SN~1006 with IXPE (see Figure~\ref{fig:fov}). The western part of the shell is proposed to interact with an atomic cloud, but the southwestern part of
the shell is expanding in a low-density medium \citep{dubner02,miceli14}. 

Another motivation for observing \src\ 
is to explore whether X-ray and radio polarization properties have large differences. For SNRs, no direct comparison of PA between X-ray and radio bands has been made.
\snr\ provides a particularly appropriate case for such an investigation, because its large angular size ($30'$) enables a detailed spatially resolved polarimetric analysis of the shell. Moreover, the high-resolution MeerKAT radio polarimetric data are available for direct comparison with the X-ray polarization data.
The X-ray and radio polarization measurements will help to reveal the magnetic field properties in the immediate and far post-shock regions, respectively, since the electrons at the X-ray bands have a short synchrotron cooling timescale. 
Radio polarization observations show that a large fraction of the magnetic fields in \snr\ lie parallel to the Galactic plane, although there is also a radial component \citep{reynoso13,cotton24}.
The origin of the two components needs further study. 

\section{Data} \label{sec:data}

\subsection{IXPE observations}
IXPE observed the southwestern shell of \snr\ (see Figure~\ref{fig:fov}) in six epochs between 2024 February 3 and March 18, providing a total exposure time of 988~ks. The photon rate in the observations was stable except for an enhanced flux in a small fraction of time due to contamination from a solar flare. We determined the good time intervals with a 3-$\sigma$ clipping for the 2--8 keV lightcurve to remove flares with anomalously high count rates. This filtered out 2/12.4/4.5~ks from the data of detector unit 1/2/3.

The software ixpeobssim \citep[vers.\ 31.0.1,][]{baldnini22} was used to reduce and analyze the IXPE data. Specifically, we used PCUBEMAP and PCUBE algorithms in the $xpbin$ command to generate polarization images and extract Stokes parameters in spatial regions, respectively. We applied the Xselect and Xspec packages (vers.\ 12.14.0b) in the HEASOFT software for spectral extraction and analysis. The spectra were weighted using the NEFF method, and the weighted response files of each region were generated using the $ixpecalcarf$ command \footnote{https://heasarc.gsfc.nasa.gov/docs/ixpe/analysis/\\ixpe\_quickstart\_v2.pdf}. The Stokes I spectra were grouped to reach at least 50 counts per bin. For the Stokes Q and U spectra, we used a constant 0.4 keV energy binning. 

To reduce the instrumental background, we adopted an energy-dependent background rejection method described in \cite{dimarco23}. Although this method removes about 40\% of the instrumental background, the remaining background still dominates over the sky X-ray background in the region of \snr. Therefore, background subtraction was conducted in the analysis for source regions.
We used the IXPE 2--4~keV band for polarization analysis throughout the paper
in order to maximize the signal to background ratio.
This band was also used for the previous study of SN1006~NE \citep{zhou2024}.

Similar to the observations of the NE shell, there was also an offset of the IXPE pointing due to difficulties in correcting imaging effects from dynamical bending of the telescope boom using only extended sources \citep{vink22}. 
IXPE detected a point source in the field of view, but its position is misaligned with that observed in the Chandra observation taken in 2012 ($\RAdot{15}{02}{03}{4}$,$\decldot{-42}{02}{19}{7}$, J2000). 
To correct the offset, we matched the positions of the point source in the IXPE and Chandra images and obtained the corrected reference coordinates of the IXPE data ($\RAdot{15}{01}{59}{3}$, $\decldot{-42}{05}{20}{0}$, J2000).

\subsection{Chandra X-ray and MeerKAT radio data}

We retrieved three archival Chandra/ACIS-I X-ray observations toward the west and southwest of SN~1006 (obs.\ IDs: 13738, 13739, and 14424), contained in the Chandra Data Collection ~\dataset[DOI: 10.25574/cdc.372]{https://doi.org/10.25574/cdc.372}. These observations were conducted between 2012 April 23 and May 4, with a total exposure of 199~ks.  We used CIAO software to reduce the data and Xspec to analyze the spectra. The main usages of the Chandra observations include image comparison with IXPE and X-ray spectral analysis. The latter constrains the absorption of X-ray emission by the foreground ISM (see Section~\ref{sec:spec} for details).

To compare the polarization between X-ray and radio bands, we used the MeerKAT L-band (856--1712\ MHz) polarization images, which have an angular resolution of $7\farcs{5}$ \citep{cotton24}.
The Meerkat data have been corrected for the intervening Galactic Faraday rotation.

\begin{figure}
\centering
	\includegraphics[width=0.5\textwidth]{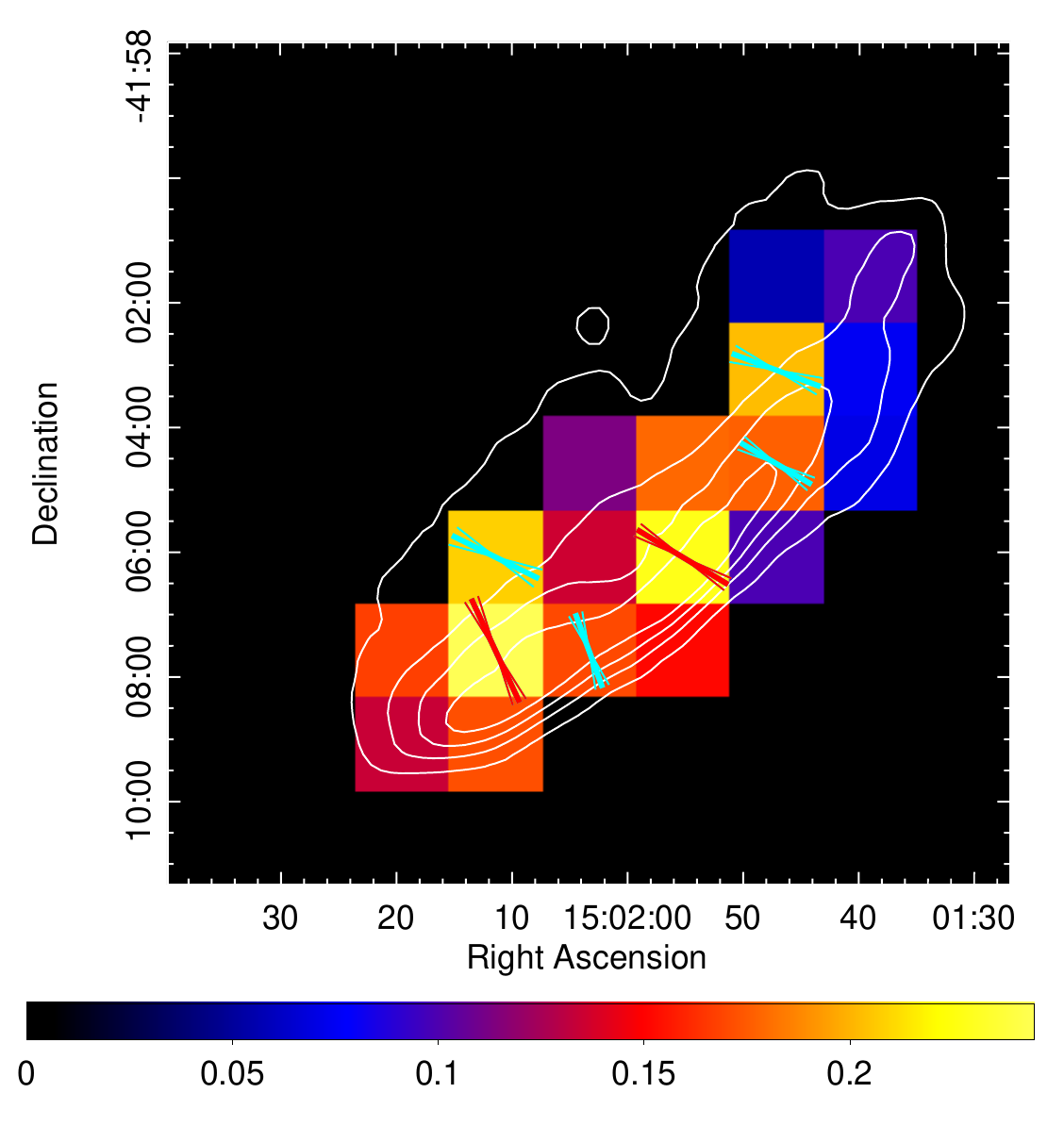}
\caption{X-ray PD overlaid with Stokes I contours and magnetic vectors (with $1\sigma$ error). Only pixels with Stokes I values larger than 175 are shown. The pixel size is $1\farcm{5}$ and the energy band is 2--4~keV. The red and cyan vectors denote the pixels with $>3\sigma$ and 2--$3\sigma$ detection, respectively. 
}
    \label{fig:pdpa}
\end{figure}

\begin{figure*}
\centering
	\includegraphics[width=\textwidth]{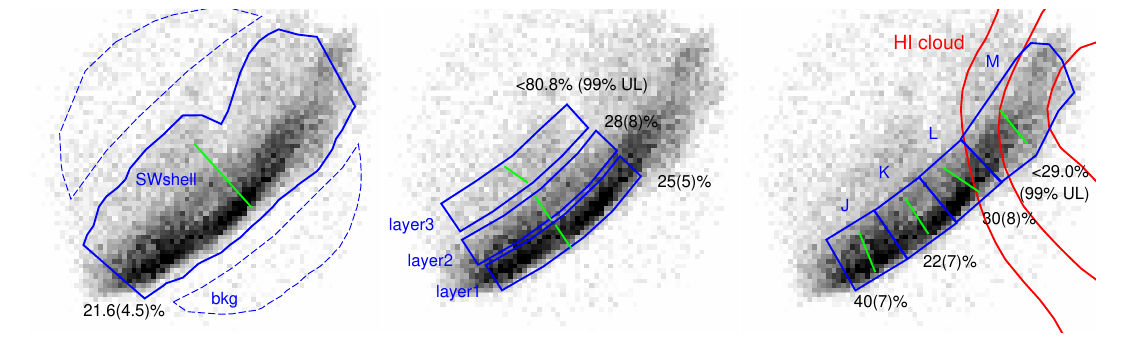}
\caption{Regions selected in our polarization analysis on top of Stokes I images. The dashed regions show the background area. The PD (with its $1\sigma$ error or 99\% upper limit) is labeled beside each region, and the mean PA values are shown with green lines (see the polarimetric results in Table~\ref{tab:pol}). The position of the HI cloud is identified by the red contours, reflecting the HI flux between 5.7--10.7~\kms. The HI data are taken from \cite{dubner02}.
}
    \label{fig:regions}
\end{figure*}

\section{Results}

\subsection{Distribution of polarization parameters} \label{sec:distribution}

Figure~\ref{fig:pdpa} shows the PD and magnetic field distribution of  \src\ rim in the 2--4 keV band for a pixel size of $1\farcm{5}$. We obtained PD and PA from the Stokes $I$, $Q$ and $U$ values in each pixel: PD$=\sqrt{Q^2+U^2}/I$, PA=0.5 arctan(U/Q) 
(measured on the plane of the sky from north to east; International Astronomical Union standard). 
Two of these pixels show PD = 23--24\%  with a $>3\sigma$ significance (red vectors, PD error of 6\%), and four pixels have PD=17--21\% with a 2--$3\sigma$ significance (cyan vectors, PD error of 6--8\%). 
The PA values range between $-20^\circ$ and $-70^\circ$ with an uncertainty of 7$^\circ$--12$^\circ$.
The magnetic vectors are nearly perpendicular to the shell, suggesting a radial magnetic field distribution. 
In the Appendix, we show the polarization maps resulting from Gaussian smoothing, which produces a consistent distribution.
We note that the significance mentioned above is a pre-trial value, not corrected for the number of pixels.
Moreover, the IXPE emission of \src\ is contaminated by a non-negligible level of instrumental background. This background needs to be removed to restore the true polarization parameters of \src.

\begin{table*}
\caption{Polarization results of eight regions by IXPE and MeerKAT observations. 
}
\centering
\begin{tabular}{l|ccc|ccc|c|c}
\hline
\hline
& \multicolumn{3}{c}{IXPE (polarimetric)}  & \multicolumn{3}{c}{IXPE (spectropolarimetric)} & Radio & \\ \cline{2-4}  \cline{5-7} \cline{8-8}
Region & PD  &  PA & $\sigma$ &  PD  &  PA &  $\Gamma$ &   PA$_r$ & PA$_r-$PA \\
& (\%) & ($^\circ$) & & (\%) & ($^\circ$) &  &  ($^\circ$) & ($^\circ$) \\ \hline
SWshell &  $21.6\pm  4.5$ & $ -48\pm 5$& 4.8 
& $21.5\pm4.7$ &  $-47\pm 6$ & $2.81\pm 0.05$
& $-28.9\pm 0.3$
& $19\pm 5$\\
layer1 &   $24.6\pm  5.3$ & $ -56\pm 6$& 4.6
& $33.2\pm 5.8$ & $-53\pm 5$ & $2.74\pm 0.06$
& $-32.6\pm 0.8$
& $23 \pm 6$ \\
layer2 &   $28.1\pm  7.7$ & $ -56\pm 7$& 3.6
& $26.9\pm 8.3$ & $-53\pm 10$ & $3.00\pm 0.09$
& $-30.2\pm 0.6$
& $26\pm 7$\\
layer3 &  $<80.8$  
&
& 2.1 
& $<75.5$  & 
& $3.16\pm 0.18$
& $-30.4\pm 0.7$ \\
J& $40.0\pm  7.0$ & $ -68\pm 4$& 5.7
& $40.0\pm 7.7$ & $-67\pm 5$ & $2.79\pm0.08$
 & $-28.2\pm 0.8$
 & $40\pm 4$\\
K&   $21.6\pm  7.3$ & $ -57\pm 9$& 3.0 
& $27.6\pm 8.0$ & $-55\pm 9$ &  $2.83\pm 0.08$
&$-30.5\pm 0.8$
& $27\pm 9$\\
L&   $29.8\pm  7.6$ & $ -34\pm 7$& 3.9 
& $35.3\pm 8.5$ & $-38\pm 7$ & $2.92\pm 0.09$ 
&$-35.8\pm 0.8$
& $-1\pm 7$\\
M & $<29.0$  
& & 2.1
& $<26.6$ & 
& $2.67\pm 0.08$
&$-21.9\pm 0.8$
\\ 
\hline
\hline
\end{tabular}
\tablecomments{The regions are denoted in Figure~\ref{fig:regions}. The polarization results are obtained using the 2--4~keV data. The errors are provided at the 1$\sigma$ confidence level, except for regions Layer3 and M with low significance, we give 99\% upper limits for the PD values. 
}
\label{tab:pol}
\end{table*}

We thus extracted X-ray polarization from regions of interest as labeled in Figure~\ref{fig:regions}, with background evaluated from the emission-free regions away from the shell. The PD, PA, and  polarization significance are listed in Table~\ref{tab:pol}. The polar plots of PD, PA, and their confidence levels are shown in the Appendix (Figure~\ref{fig:polar}).

A large background region is necessary to enhance the statistical quality of polarization parameters. Therefore, we used a large background area ``bkg'' that covers two regions in the SW and NE (in the SNR interior). Weak thermal emission exists in the SNR interior below 2~keV \citep{li16}, but our polarization analysis focuses on 2--4 keV. Moreover, we found that the NE portion of ``bkg'' (see Figure~\ref{fig:regions}) is only $\sim 13\%$ brighter than in the SW in 2--4 keV, meaning that SNR contribution is subdominant in the ``bkg'' NE region.
The X-ray emission from the ``bkg'' region is consistent with being unpolarized with  PD(bkg)$<9.2\%$ at 99\% confidence level. 

We selected ``SWshell'' for the entire SW shell, including both the bright rim and the faint part. We detected polarization in this region at a $4.8\sigma$ level and obtained PD=$21.6\%\pm 4.5\%$ and PA=$-48^\circ\pm5^\circ$. These numbers are close to the values in the northeastern shell: PD(NE)=$22.4\%\pm 3.5\%$ and PA(NE)=$-45^\circ\pm 4^\circ$ \citep{zhou2024}. 

To explore the polarization variation along the radial direction, we defined regions ``layer1'', ``layer2'', and ``layer3'' from the bright rim to the areas behind the shock.
The former two regions have PD values of $26\%\pm 5\%$, $28\%\pm 8\%$ (1-$\sigma$ error), respectively, while the 99\% upper limit of PD in ``layer3'' is 80.8\%.
Considering the error range, the current data do not reveal a significant change in polarization along the radial direction.

We found a variation of PD and PA along the azimuthal regions ``J'', ``K'', ``L'', and ``M''. The PD is largest in ``J'' (PD(J)=$40\%\pm 7\%$), with a high significance of 5.7$\sigma$, but reduced at its neighboring region ``K'' (PD(J)=$22\%\pm 7\%$)). 
``M'' corresponds to the region impacted by an HI cloud, where PD(M)$< 29.0\%$ at a confidence level of 99\%.
Moreover, the PA presents an increasing trend from PA(J)=$-68^\circ\pm 4^\circ$ to PA(L)=$-34^\circ\pm 7^\circ$. The difference in PA is significant between these two regions, at a 4.2$\sigma$ level, suggesting that PA in \snr\ is not perfectly parallel, but varies along the azimuthal angles.

To verify whether PA in \src\ aligns along the shell direction, we reconstruct the event files based on the circular polarization geometry model using the  $xpstokesalign$ command in $ixpeobssim$. All the Stokes Q and U parameters of each event are rotated with respect to the SNR  center at R.A.~(J2000)$=\RAdot{15}{02}{51}{7}$, Decl.\ (J2000)=$\decl{-41}{56}{33}$ \citep{reynolds86,winkler97}, so that we would find the reconstructed PA of $0^\circ$ if the initial PA aligns with the shell. With the aligned data, we found PA$_{\rm align}$ of $0^\circ\pm 5^\circ$ in region ``SWshell'', and $-11^\circ\pm 5^\circ$, $-9^\circ\pm 9^\circ$, $-6^\circ\pm 7^\circ$ and $-19^\circ\pm 14^\circ$ for regions ``J'', ``K'', ``L'', and ``M'', respectively.
This supports the finding that PA in \src\ is tangential to the shell, and the magnetic fields are radially distributed.

\subsection{Spectro-polarimetric analysis} \label{sec:spec}

\begin{figure}
\centering
	\includegraphics[width=0.5\textwidth]{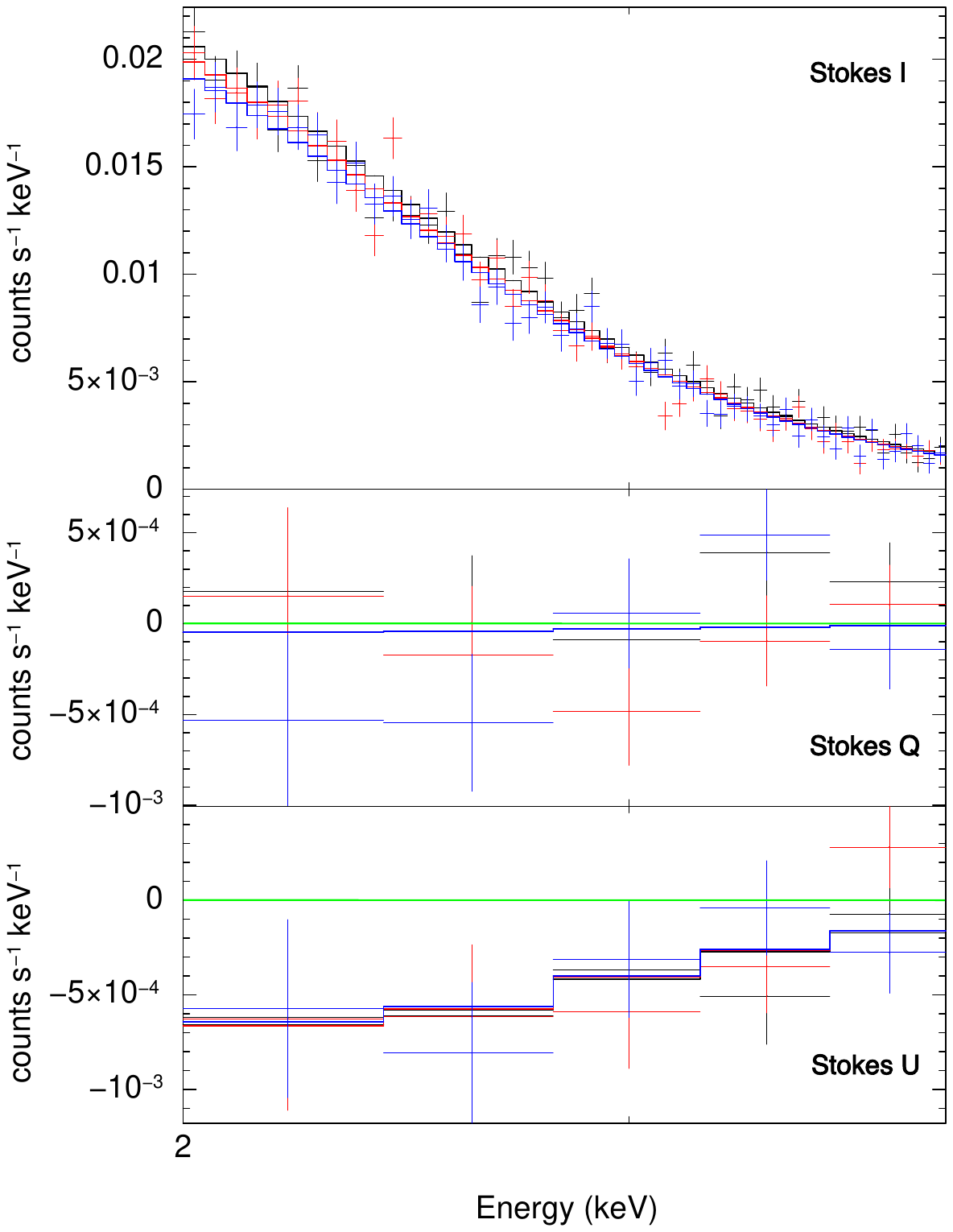}
\caption{Stokes I, Q, and U spectra of region ``SWshell'' fitted using the $tbabs*polconst*powerlaw$ model. The black, red, and blue data are extracted from DU 1, 2, and 3, respectively.
}
    \label{fig:spec}
\end{figure}

We performed spectro-polarimetric analysis for the regions as shown in Figure~\ref{fig:regions}.  This analysis has some advantages compared to the pure polarimetric method (PCUBE). Firstly, it allows simultaneous measurements of spectral and polarization parameters, with a correction for the vignetting effect. Moreover, it can provide the (posterior) probability distribution of PD and PA, and thus allows a more reliable comparison of PD or PA between regions.

We extracted weighted Stokes I, Q, and U spectra from these regions and subtracted the background Stokes spectra. These spectra in 2--4~keV were then fitted using an absorbed power-law model with a constant polarization: $tbabs*polconst*powerlaw$ in Xspec (see Figure~\ref{fig:spec} for region ``SWspec''). We adopted solar abundance from \cite{asplund09} and the Tuebingen-Boulder ISM absorption model $tbabs$ which calculates the cross section for X-ray absorption by muti-phase gas and dust.  
The spectra of \snr\ in the IXPE 2--4~keV energy band are not sensitive to the absorption parameter, i.e., equivalent atomic column density $N_{\rm H}$, 
because spectra above 2 keV are not affected by the moderate column densities toward SN 1006.
Therefore, we fitted 1--8~keV Chandra X-ray spectrum in ``SWshell'' region with the $tbabs*powerlaw$ model and obtained  $N_{\rm H}=3.27\pm 0.02\times 10^{21}~\cm^{-3}$. This best-fit value is then used and fixed for the IXPE spectral fitting. 

We applied the Markov Chain Monte Carlo (MCMC) method \citep{mackay03} to provide the mean values and error ranges of parameters. 
We use the Goodman-Weare algorithm \citep{goodman10} with 20 walkers and a total length of 200,000, with the first 100,000 steps discarded.
After reaching equilibrium, the MCMC chains can be converted to posterior probability distributions of parameters, and thus provide parameter error ranges as tabulated in Table~\ref{tab:pol}.
The PD and PA values from this spectro-polarimetric approach are consistent with the polarimetric results within the error range.

The spectro-polarimetric results demonstrate a variation of X-ray PD along the shell: The MCMC chains provide the posterior probability distributions of PDs in regions and the combined region ``JKL'' (PD(JKL)=$31.7\%\pm 4.6\%$; see Figures~\ref{fig:corner} and \ref{fig:compare} in the Appendix). The probability is 99.56\% and 99.83\% for PD(JKL)$>$PD(M) and PD(J)$>$PD(M), respectively.
Using $\chi^2$ statistics to fit the spectra results in a consistent probability of $99.41\%$ for PD(JKL)$>$PD(M) (see the Appendix).

\subsection{Radio polarization angle}

Radio and X-ray synchrotron emissions are radiated by electrons of different energies and lifetimes, with radio emission arising typically from $10^8$--$10^{10}$~eV electrons, and X-ray synchrotron emission from typically $10^{13}$--$10^{14}$~eV electrons. 
Therefore, it is of interest to investigate whether the polarization properties of these two bands are the same.
In this paper, we only compare PA values, 
given that the missing flux issue in the radio interferometric observations with MeerKAT makes a reliable radio PD measurement difficult \citep[see][for detailed explanation]{cotton24}.
From the MeerKAT L-band polarization data, we obtained PA values in the eight regions and listed them in Table~\ref{tab:pol}.
The MeerKAT data have a finer angular resolution of $7\farcs{5}$ than  IXPE.
To compare these two data, we first smooth the MeerKAT polarization images to reach an angular resolution of $30''$.

The radio PA values (PA$_r$) in each region are calculated using the polarized flux (PF) and PA images:
\begin{equation}
\begin{split}
   \rm{ PA}_r & =0.5 \arctan \frac{\sum U_i}{\sum Q_i} \\
   & =0.5 \arctan {\rm \frac{\sum PF_{\it i}  \sin (2PA_{\it i})}{\sum PF_{\it i} \cos (2PA_{\it i})}}
\end{split}
\end{equation}

where the subscription $i$ means the $i$th pixel.

The radio PA$_r$ values in all regions are around $-30^\circ$ (see Table~\ref{tab:pol}), except for the region ``M'' that has a slightly larger PA$_r$ of $-22^\circ$.
This deviates significantly from the X-ray PA results.
At the region ``SWshell'', the radio PA$_r=-28.9^\circ\pm 0.3^\circ$ differs from the X-ray PA$=-48^\circ\pm 5^\circ$ at the $3.8\sigma$ level. The discrepancy is most prominent in the region ``J'', which has PA$-$PA$_r\approx 40^\circ$ ($9.8\sigma$).
Therefore, the X-ray and radio polarization, or the magnetic fields traced by these two bands, have large differences.

\section{Discussion}

\subsection{A variation of X-ray PD along the shell}

We have found a significant variation of X-ray PD along \src\ (see Sections~\ref{sec:distribution} and \ref{sec:spec}), with a peak PD of $40\%\pm 8\%$ in the southern region ``J'' and a low PD of $<27\%$ ($99\%$ upper limit) in the western region ``M'' (see the regions in Figure~\ref{fig:regions}).
This suggests that the magnetic field is quite ordered in the eastern part of the SW shell (high PD), but is more randomly distributed in the west (low PD), where the SNR is suggested to interact with an atomic cloud \citep[]{dubner02, miceli14}.
The cloud spatially corresponds to a X-ray-dim area (see Figure~\ref{fig:fov}), which can be attributed to a lower cutoff
energy of the X-ray synchrotron emission in that region \citep{miceli14}.
\cite{miceli14} estimated the atomic gas density to be 9--16~$\cm^{-3}$ from the HI observations, and the cloud must be partially in front of SN1006, as the X-ray absorption
column is larger in that region. One might expect enhanced thermal X-ray emission
in those regions in which the shell interacts with the dense cloud. 
This is not observed, which puts some constraint on the density of the postshock gas,
which depends on the assumed electron temperature. 
\cite{miceli14} estimated post-shock gas densities of $<12, 3, 0.3~\cm^{-3}$,  for electron temperatures of $0.05, 0.25, 2$~keV, respectively. Since the shell is at the edge
of the atomic cloud, the local density may be lower than the 9-16~$\cm^{-3}$ inferred
for the atomic cloud. But the postshock density must be higher than around the rest of the shell of SN1006, which has been estimated to be 0.16--0.34~$\cm^{-3}$ \citep{acero07,katsuda09,miceli14,giuffrida22}.

In region ``J'' with fairly ordered magnetic fields, 
the magnetic turbulence scale $l_{\Delta B}$ is inferred from the scale of its X-ray structure.
Since the X-ray photons mainly come from the bright rim, the magnetic turbulence scale is likely similar to or slightly smaller than the rim thickness  \citep[$11''$][]{ressler14}, $\sim 0.1$~pc at region ``J'' for an SNR distance of 2.18~kpc \citep{winkler03}.
At region ``M'', the lower PD indicates a depolarization due to a smaller turbulence scale. 

Here we discuss two possible scenarios that may explain the decreased PD at the shock-cloud interaction region: instability caused by density fluctuation of the ISM and instability induced by CRs. 

First, when a shock propagates through an inhomogeneous ISM, it triggers plasma instability, leading to magnetic turbulence with a characteristic scale proportional to the length scale of the density dispersion \citep{inoue13,xu17}.
This mechanism is expected to occur in Cas~A, which shows a low X-ray PD in the rim \citep[$4.5\pm 1.0\%$;][]{vink22}, since this young SNR is evolving in a highly clumpy circumstellar medium \citep{vink96,chevalier03,koo23}. Tycho, with a PD of $12\pm 2\%$  \citep{ferrazzoli23}, is also likely to experience some level of magnetic turbulence due to its nonuniform ambient medium \citep{williams13,zhou16,tanaka21}.
In \src, the density of the atomic cloud close to region ``M'' is two orders of magnitudes larger than the average value near SN~1006 \citep[0.04--0.85~$\cm^{-3}$,][]{acero07,katsuda09,giuffrida22}, providing an inhomogeneous condition that may generate small-scale magnetic turbulence.
In contrast, the PD reaches 40\% at the southern region ``J'', where the shell is expanding into a low-density medium. Therefore, density fluctuation is unlikely to play a significant role in driving magnetic turbulence at region ``J''. 

The alternative explanation involves the streaming instability of CR currents, which causes strong amplification and turbulence of magnetic fields in the shock upstream \citep{bell04}. This instability also generates plasma fluctuations and thus ripples the shock surface, triggering plasma and magnetic turbulence downstream \citep{caprioli14,bykov24}.    

\begin{figure}
\centering
	\includegraphics[width=0.5\textwidth]{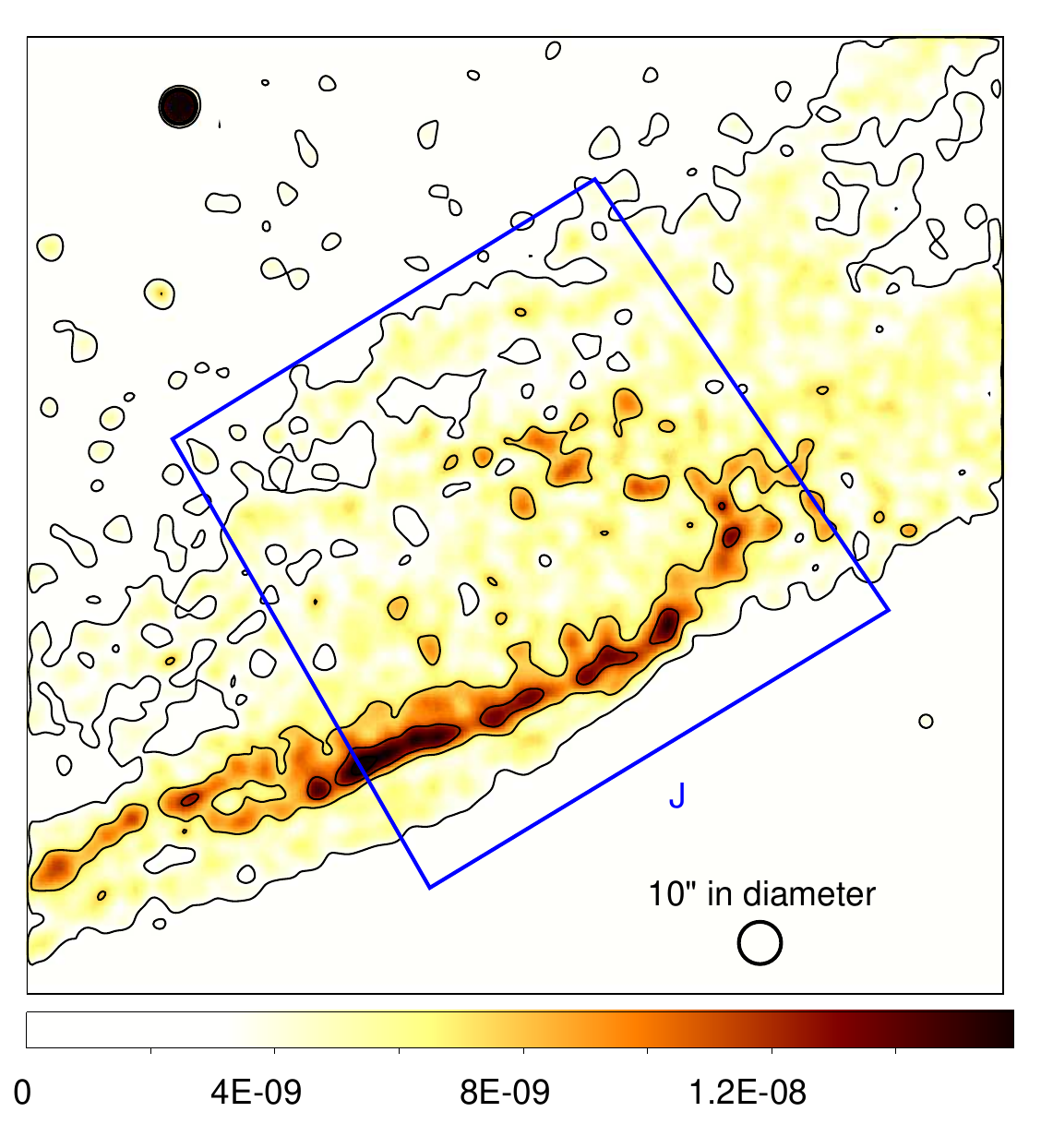}
\caption{Clumpy structures of the synchrotron emission in the southwestern shell of SN 1006 revealed with Chandra 2--8 keV flux image (in the unit of count~$\ps \cm^{-2}$). The image with a pixel size of $0\farcs{49}$ is smoothed with a Gaussian kernel ($\sigma$) of 4 pixels. The blue box shows the region ``J'' and the black circle marks a scale of $10''$ ($\sim 0.1$~pc).
}
    \label{fig:chandra}
\end{figure}

The length scale of Bell instability in the preshock inversely scales with the ambient density $l_{\rm Bell}\propto n_0^{-1} V_s^{-3} E_{\rm max} B_0 \propto n_0^{-1}$, where $V_s$, $E_{\rm max}$ and $B_0$ are the shock speed, maximum energy of the CRs and the ambient magnetic strength. 
Assuming a similar turbulence scale between the preshock and postshock, we estimate $l_{\rm Bell}\sim l_{\Delta B} \sim 0.1$~pc at region ``J''.  This would indicate a nonuniform X-ray brightness at this scale,  which is found in the Chandra X-ray image as shown in Figure~\ref{fig:chandra}.
In the high-density region with an atomic cloud, the Bell instability occurs on a smaller scale.
For example, assuming $n_0=0.05~\cm^{-3}$ near region ``J'' and $=0.5~\cm^{-3}$ near region ``M'' \cite[cloud density is 9--16$\cm^{-3}$][]{miceli14}, $l_{\rm Bell}$ is expected to be 10 times smaller in region ``M''. 
Therefore, the depolarization can explain the lower PD found in the region ``M''. 
We note that the shock velocity, $E_{\rm max}$, and $B_0$ may also be different between regions in \src, but these values are unknown so far.

\begin{figure}
\centering
	\includegraphics[width=0.5\textwidth]{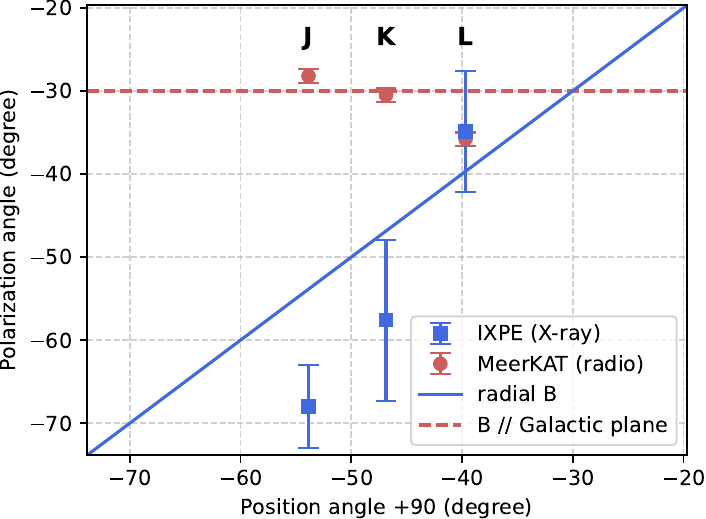}
\caption{Polarization angles measured in 3 regions with IXPE and MeerKAT compared to the position angles ($+90^\circ$, with $0^\circ$ to the north) of the regions. The solid and dashed lines represent two magnetic configurations, radial magnetic orientation and magnetic fields parallel to the Galactic plane, respectively. The error bars show the $1\sigma$ range.
}
    \label{fig:pa}
\end{figure}

\subsection{Large difference between X-ray and radio PA}

The large discrepancy between the X-ray and radio PA (see Table~\ref{tab:pol}) constitutes strong evidence that the magnetic fields traced by these two bands are not the same. 
We notice that the magnetic orientation is predominantly radial as revealed by the X-ray polarization, while consistent with the direction of the Galactic plane ($\sim 60^\circ$, for PA$_r\sim -30^\circ$) as probed by the radio emission (see Figure~\ref{fig:pa}).

As shown in Figure~\ref{fig:pa}, X-ray PA values increase with the position angle of the regions from $-144^\circ\pm 4^\circ$ in the region ``J'' to $-130^\circ\pm 5^\circ$ in region ``L'' (with $0^\circ$ to the north). The difference between PA and position angle $+90^{\circ}$ is within $\sim 14^\circ$ for the three regions with significant polarization detection.
This correlation supports the idea that X-ray emission is generated in an area with predominantly radial magnetic fields. There is also a tiny deviation from the perfectly radial distribution, which may be real or due to the random error of the measurement.
Moreover, the overall X-ray PA in SN~1006~SW and SN~1006~NE can be described with a tangential model, with PA$_{\rm align}~({\rm SWshell}) =0^\circ\pm 5^\circ$ (see Section~\ref{sec:distribution}) and PA$_{\rm align}~({\rm NEshell}) =1^\circ\pm 5^\circ$ \citep{zhou2024}.

We also calculated the MeerKAT PA for the northeastern shell of SN~1006: PA$_r=-35.6^\circ\pm 0.8^\circ $, $-32.2^\circ\pm 0.8^\circ $ and $-27.0^\circ\pm 0.7^\circ $ for regions ``A'', ``B'', and ``C'' \citep[see regions in Figure~3 of][]{zhou2024}, respectively. The corresponding IXPE X-ray PA values are $-49.5^\circ\pm 6.2^\circ $ and $-43.7^\circ\pm 6.2^\circ$, and $-40.0^\circ\pm 8.3^\circ$. Using the $\chi^2$ test ($\chi^2=\sum_i \frac{[\rm PA({\it i})-PA_r ({\it i})]^2}{\rm Var[PA({\it i})-PA_r[{\it i})]}=10.76$,  $i$ for regions ``A'', ``B'', and ``C''), the X-ray and radio PA values are different at the 99\% confidence level, less significant than in \src.

The radial magnetic field orientation in \snr\ revealed with IXPE is inconsistent with the compression origin of the magnetic fields. This is because shock compression of the interstellar magnetic fields can amplify the tangential component in the post-shock, while the radial component remains unaffected, resulting in tangentially distributed magnetic fields. The radially distributed magnetic fields have been observed in two younger SNRs, Cas~A and Tycho, and the northeastern shell of SN~1006. An explanation of this orientation is the stretching of turbulent magnetic fields along the shock direction \citep[see][for details]{vink22,ferrazzoli23,zhou2024}.

Hereafter, we discuss the physical origins of the different magnetic field configurations revealed by X-ray and radio polarimetry. Firstly, the X-ray and radio emission comes from different regions.
The charateristic electron cooling timescale in the magnetic field $B$ is $\tau_{\rm cool}=634/(B^2E)~\s$ \citep[e.g.,][]{reynolds98, vink20}. The very-high-energy electrons that produce X-ray emission cool faster, and thus, the X-ray emission traces regions closer to the particle acceleration sites, while the lower-energy electrons that give rise to the radio emission would have a broader distribution in the post-shock region. 
The width of the X-ray synchrotron filament is determined by the advection length scale of the cooling electrons \citep{vink20}: $l_{\rm adv}=\tau_{\rm cool} (V_{\rm s}/4)\approx 0.25\left( \frac{h\nu}{3~{\rm keV}}\right)^{-1/2} \left( \frac{B}{30~{\rm \mu G}}\right)^{-1/2} \left( \frac{V_{\rm s}}{5000~\km\ps }\right) \rm pc$, with $B\sim 30~\mu$G \citep{acero10,xing16} and shock velocity $V_{\rm s}\sim 5000~\km\ps$ \citep{winkler14}. This corresponds to an angular width of $24''$, similar to the observed width of the X-ray filament in the SW shell.
The radio synchrotron structures are more extended than the X-ray and have a larger depth along the line-of-sight.

Besides the electron distribution, the magnetic fields in the immediate post-shock and the far post-shock can also be different.
The strong amplification of the turbulent magnetic fields happens at the immediate post-shock and decays in the far post-shock \citep{bykov24}.
In the Bell instability scenario, the highly amplified magnetic fields downstream contain two layers, with a thin layer of compressed transverse magnetic fields and followed by a wider layer of predominantly radial magnetic fields generated by a turbulent dynamo mechanism \citep{bykov24}. An alternative possibility is that small-scale, strongly magnetized structures (such as filaments) that are unresolved at Chandra's resolution contribute to the X-ray emission.
On the other hand, the radio emission is from both small-scale structures and more extended regions. Note that the synchrotron emissivity scales with magnetic field strength
as $\propto B^{1+\Gamma}$. Since the radio spectral index is smaller 
\citep[$\Gamma=\alpha+1\approx 1.6$, e.g.,][]{reynoso13} than in X-rays ($\Gamma\approx 2.8$), the X-ray emission is much more biased toward plasma with relatively large magnetic-field strengths.

We have discussed two possible aspects that may lead to the PA difference: 1) X-rays come from regions closer to the shock than the radio; 2) X-rays come from regions with stronger magnetic fields. Aspect 1) alone has the problem of why the magnetic field in the far downstream would turn back to the Galactic value. We consider it likely that both of these effects contribute to the observed differences between the radio and X-ray PA values.

Therefore, our X-ray polarization measurement supports the hypothesis that the magnetic fields have a turbulent origin rather than a compression origin. 
The X-ray-probed magnetic component is a newly generated component whose energy is gained from the very-high-energy particles that are freshly accelerated.
The coincidence between the radio-probed magnetic fields and the Galactic plane is also interesting, as it may suggest that CR-driven turbulence dissipates far downstream, leaving the Galactic background field dominant.

\section{Conclusion}

IXPE has observed both the NE \citep{zhou2024} and SW shell of \snr. This paper reports on the X-ray polarization properties in the SW shell. The X-ray polarization degree and angle are $21.6\%\pm 4.5\%$ and $-48^\circ\pm 5^\circ$ in \src, similar to that in the NE, thus supporting an overall symmetry for the double limbs of the SNR.

Nevertheless, the new IXPE observations toward the SW shell reveal a variation of polarization properties along the shell. We detected a change in PD, which peaks in the south ($40\%\pm 7\%$) and decreases below $\sim 27\%$ (99\% upper limit) in the western shell impacted by a denser ambient medium. This reflects a correlation between enhanced magnetic turbulence and ambient gas density.
We interpret that the decreased PD at region ``M'' results from either ISM-fluctuation-caused MHD instability or a smaller length scale of Bell instability.

The X-ray polarization angle also varies along the shell and largely deviates from the radio polarization angle. The magnetic fields probed by the X-ray band tend to follow a radial direction (along the shock normal) while those probed by the radio band follow the Galactic plane direction. We suggest that the X-ray polarization probes freshly amplified magnetic fields from small-scale structures in immediate postshock, while the radio polarization is contributed by more extended regions. 
These results highlight the importance of X-ray polarimetry observations in studying magnetic fields and CR acceleration in SNRs.

\vspace{0.3in}

We thank Dr.\ Gloria Dubner for kindly providing the HI data for SN~1006.
We also thank Dawoon E.\ Kim for providing the scripts to plot polar plots of polarization parameters.
The Imaging X-ray Polarimetry Explorer (IXPE) is a joint
US and Italian mission. The US contribution is supported by
the National Aeronautics and Space Administration (NASA)
and led and managed by its Marshall Space Flight Center
(MSFC), with industry partner Ball Aerospace (contract
NNM15AA18C). The Italian contribution is supported by the
Italian Space Agency (Agenzia Spaziale Italiana, ASI) through
contract ASI-OHBI-2017-12-I.0, agreements ASI-INAF-2017-
12-H0 and ASI-INFN-2017.13-H0, and its Space Science Data
Center (SSDC) and by the Istituto Nazionale di Astrofisica
(INAF) and the Istituto Nazionale di Fisica Nucleare (INFN) in
Italy. This research used data products provided by the IXPE
Team (MSFC, SSDC, INAF, and INFN) and distributed with
additional software tools by the High-Energy Astrophysics
Science Archive Research Center (HEASARC) at NASA
Goddard Space Flight Center (GSFC). 
P.Z. acknowledges the
support from NSFC grant No.\ 12273010.
P.S. acknowledges partial support from NASA Contract NAS8-03060.
R.F., E.C, and P. So. are partially supported by MAECI with grant CN24GR08 “GRBAXP: Guangxi-Rome Bilateral Agreement for X-ray Polarimetry in Astrophysics”.  N.B. was supported by the INAF MiniGrant ``PWNnumpol - Numerical Studies of Pulsar Wind Nebulae in The Light of IXPE."
C.-Y.N. is supported by a GRF grant of the Hong Kong Government under HKU 17304524.
E.M.R. is member of the Carrera del Investigador Cient\'\i fico of CONICET (Argentina) and is partially funded by
CONICET grant PIP 11220220100211.


%

\vspace{5mm}
\facilities{IXPE, MeerKAT, Chandra}


\software{ixpeobssim \citep[vers.\ 31.0.1]{baldnini22}, DS9 \citep{joye03,sao00}, Xspec \cite[vers.12.14.1][]{arnaud96}, Astropy \citep{astropy:2013, astropy:2018, astropy:2022}, emcee \citep{foreman-mackey12}}





\appendix

\begin{figure}
\centering
	\includegraphics[width=0.5\textwidth]{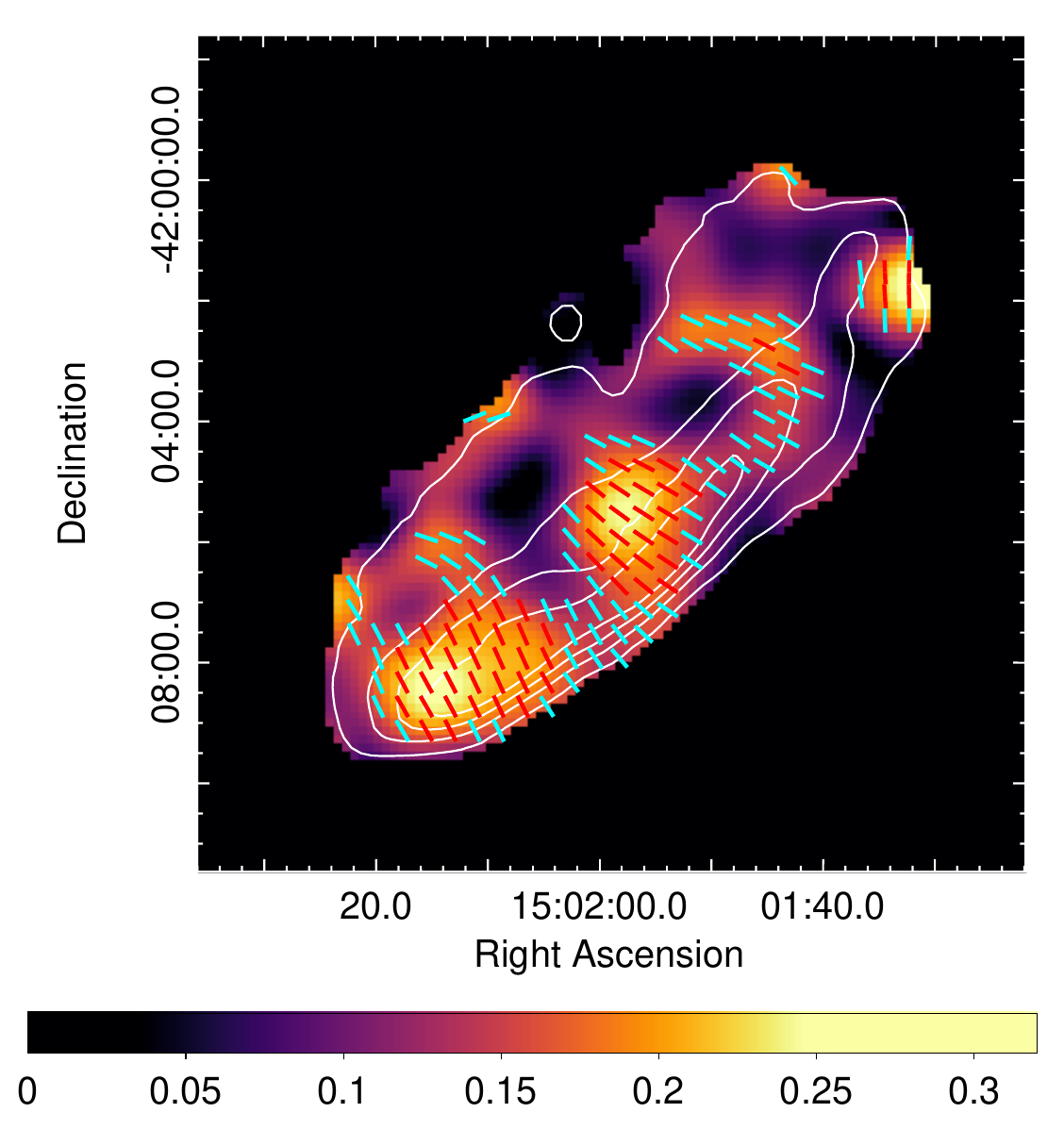}
\caption{X-ray PD image resulting from smoothing the Stokes I, Q, and U 
with a Gaussian kernel of width $\sigma=36''$. The Stokes I contours and magnetic vectors are overlaid. The significance
is determined as explained in \cite{prokhorov24}.
Only for significant polarization detections are the vectors
overlaid: red for over $3~\sigma$ pixels and cyan for 2--$3~\sigma$ pixels.
}
    \label{fig:pd_smooth}
\end{figure}

\section{Polarization map with Gaussian smoothing}

We applied a smoothing technique to produce the polarization map to improve the image statistics \citep{prokhorov24}. We first produce Stokes I, Q, and U and their variance maps with a pixel size of $8''$. Then, Stokes I, Q and U images are smoothed with a Gaussian kernel of width $\sigma=36''$ to produce PD and PA images as shown in Figure~\ref{fig:pd_smooth}. 
The variances are smoothed with the kernel squared, and
can be used to obtain polarization statistics per pixel, 
as $\chi^2_{2,i}=Q_i^2/{{\rm Var}(Q_i)} + U_i^2/{{\rm Var}(U_i)}$.

There are two ``hot spots" along the rim with PD$\sim 25\%$,
and peak significances of $5.2\sigma$ ($\chi^2_2=30.4$) and
 $5.0\sigma$ ($\chi^2_2=28.9$). The positions and values are similar to those in Figure~\ref{fig:pdpa}, which is produced using a binning method (bin size of $1\farcm{5}$). Therefore, the smoothing and binning methods provide consistent results.

\begin{figure}
\centering
\includegraphics[width=\textwidth]{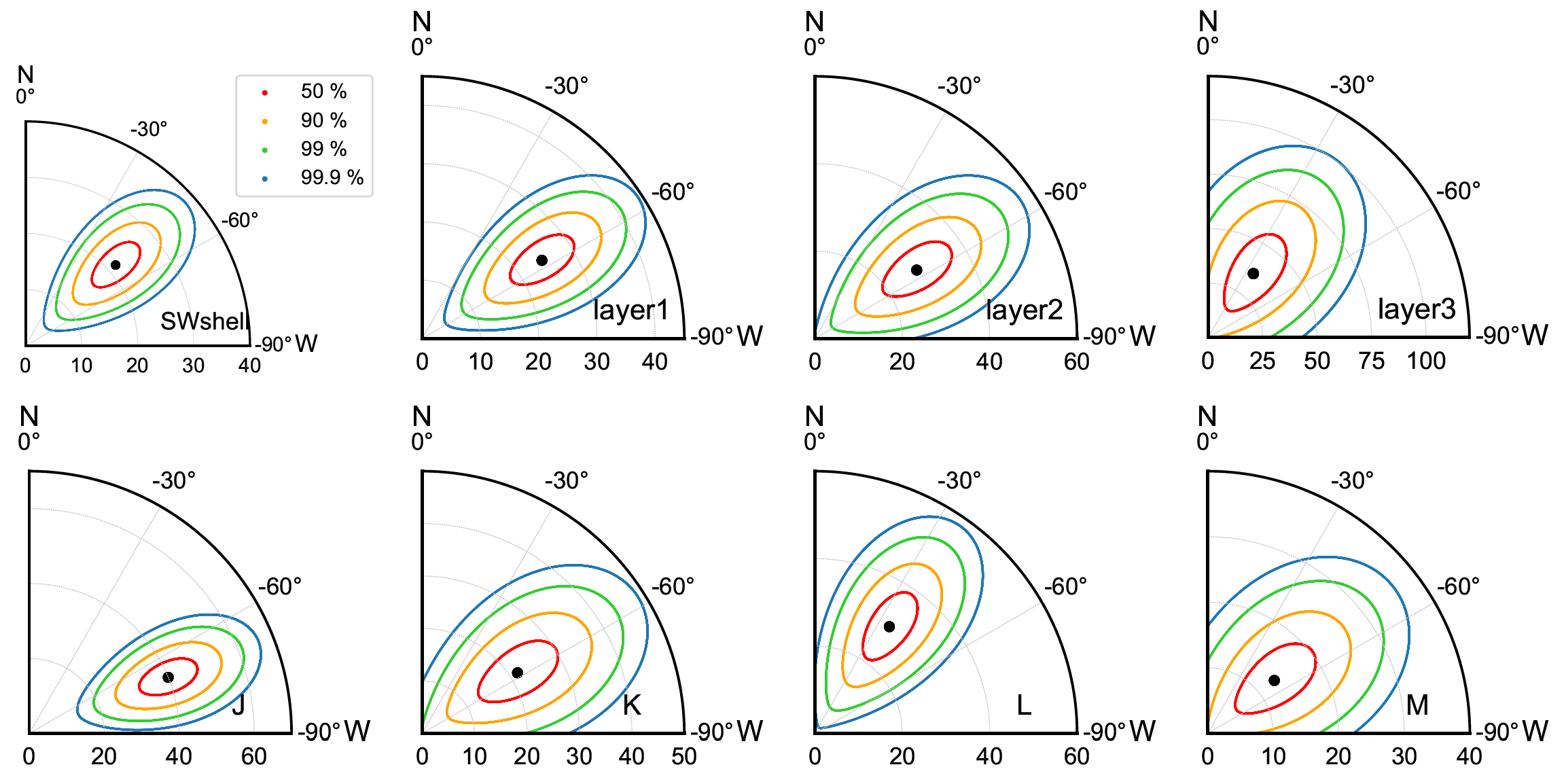}
\caption{Polar plots of polarization parameters for regions in the 2–4 keV band (see regions in Figure~\ref{fig:regions}).
Each polar plot shows the PD (\%) and PA in radial and position angle coordinates, respectively. The best-fit PD and PA values are denoted by black dots. 
}
    \label{fig:polar}
\end{figure}

\begin{figure}
\centering
\includegraphics[width=0.48\textwidth]{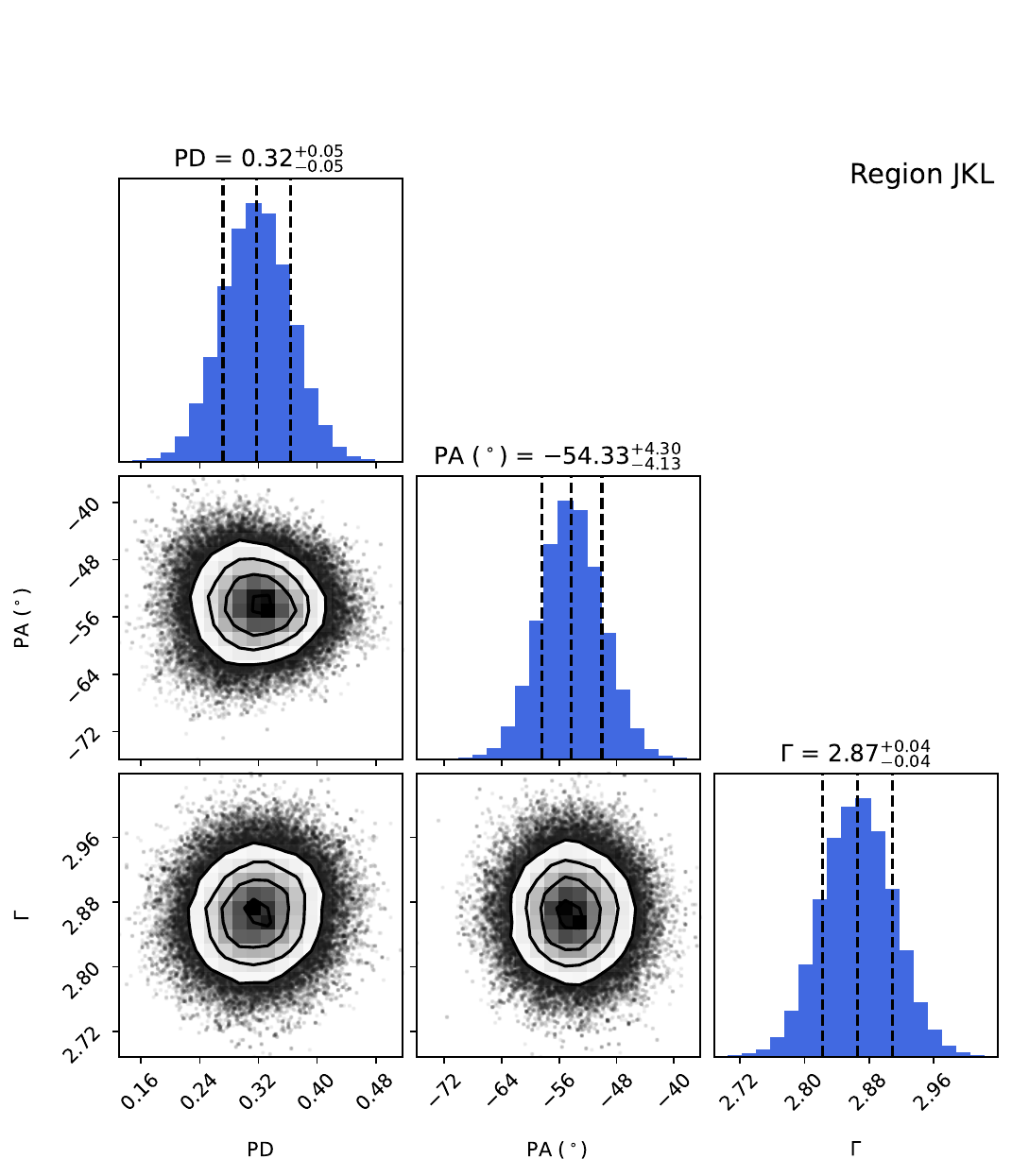}
\includegraphics[width=0.48\textwidth]{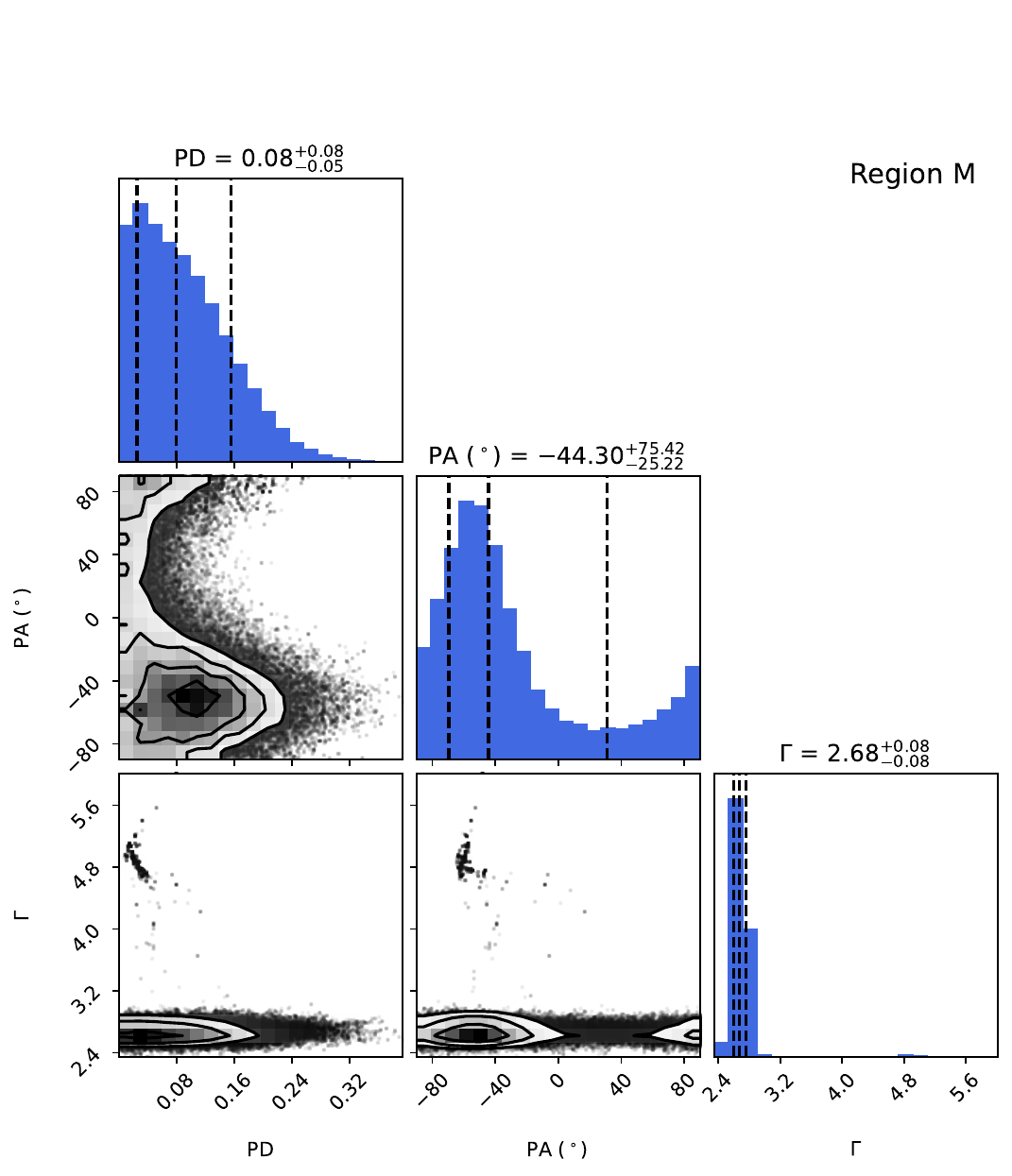}
\caption{Two-dimensional correlation distributions of parameters in the spectro-polarimetric analysis for regions JKL (left) and M (right). The contours provide the 0.5, 1, 1.5, $2 \sigma$ equivalent range. The histograms show the posterior distribution of PD, PA, and spectral index $\Gamma$, and the vertical lines denote the 0.16, 0.5, and 0.84 quantiles of the samples. 
}
    \label{fig:corner}
\end{figure}

\begin{figure}
\centering
\includegraphics[width=0.5\textwidth]{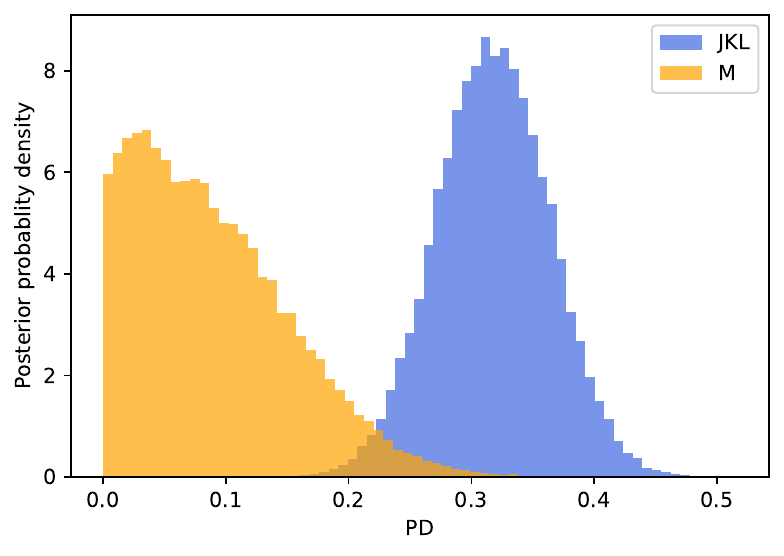}
\caption{Posterior distribution of PD in regions ``JKL'' and ``M''.
}
    \label{fig:compare}
\end{figure}

\section{Comparison of PD between regions using MCMC analysis and $\chi^2$ statistics}

The MCMC analysis of spectro-polarimetric data provides the posterior probability distribution for parameters in region ``M'' and the combined region ``JKL''.
Figure~\ref{fig:corner} shows the two-dimensional correlation between parameters and the posterior distributions of individual parameters.
In Figure~\ref{fig:compare}, we compare the PD posterior probability distribution between region ``JKL'' and ``M''. The probability of PD(M)$< $PD(JKL) is derived as 99.56\%.

In addition to the MCMC analysis, we also used the  $\chi^2$ statistic for the spectral fit to compare PD between regions ``JKL'' and ``M''. Using the {\it steppar} command, we calculated the $\chi^2$ statistic value as a function of PD from 0.02 to 0.5 with a step of 0.02. The  $\Delta \chi^2=\chi^2-\min(\chi^2)$ distribution can be converted to a probability distribution of PD. The probability of PD(M)$<$PD(JKL) is derived as $1-\sum_{i=0}^{25} [\sum_{j=0}^i p_{\rm JKL}(j)]p_{\rm M}(i)=$99.41\%, where $p_{\rm JKL}(i)$ is the probability of the real PD(JKL) value falling in the $i$th interval (intervals 0, 1, 24,  and 25 correspond to PD=0--0.02, 0.02--0.04, 0.48--0.5, and 0.5--1, respectively). This probability is consistent with that obtained from the MCMC method.

\end{document}